\documentclass{iopart}

\usepackage{geometry}
\usepackage{caption}
\usepackage{subcaption}
\usepackage{xspace}
\expandafter\let\csname equation*\endcsname\relax
\expandafter\let\csname endequation*\endcsname\relax
\usepackage{amsmath}
\usepackage{amssymb}
\usepackage{physics}
\usepackage{isotope}
\usepackage{hyperref}
\usepackage[nameinlink]{cleveref}
\usepackage[numbers,sort&compress]{natbib}
\usepackage{orcidlink}

\crefname{table}{Tab.}{Tabs.}
\Crefname{table}{Table}{Tables}
\crefname{figure}{Fig.}{Figs.}
\Crefname{figure}{Figure}{Figures}
\crefname{equation}{Eq.}{Eqs.}
\Crefname{equation}{Equation}{Equations}

\makeatletter

\renewcommand\@fnsymbol[1]{\arabic{#1}}

\renewcommand\@makefnmark{%
  \hbox{\textsuperscript{\normalfont\@thefnmark}}%
}

\renewcommand\@makefntext[1]{%
  \parindent 1em%
  \noindent
  \hb@xt@0.4em{\hss\textsuperscript{\normalfont\@thefnmark}}%
  #1%
}

\makeatother

\DeclareMathAlphabet{\curly}{OMS}{cmsy}{m}{n}

\newcommand*{\elem}[2]{\ensuremath{\isotope[#2]{\mathrm{#1}}}}

\newcommand*{\hw}{\ensuremath{{\hbar\Omega}}\xspace}
\newcommand*{\Nmax}{\ensuremath{{N_\mathrm{max}}}\xspace}

\begin{document}


\title[High-precision ab initio nuclear theory: Learning to overcome model-space limitations]{High-precision ab initio nuclear theory:\\Learning to overcome model-space limitations}

\author{Marco Kn\"{o}ll\orcidlink{0009-0001-6023-8657}}
\vspace{5pt}
\address{Institut f\"ur Kernphysik, Technische Universit\"at Darmstadt, 64289 Darmstadt, Germany}
\vspace{5pt}
\ead{{\normalfont marco.knoell@physik.tu-darmstadt.de}}


\begin{abstract}
\noindent High-precision predictions of nuclear properties are a central objective of ab initio nuclear structure theory.
However, state-of-the-art many-body methods rely on truncated model spaces to render the nuclear many-body problem tractable, which remains a major source of theoretical error in computations of nuclear observables. 
In recent years, machine learning, and artificial neural network approaches in particular, have emerged as a powerful data-driven framework for learning convergence patterns directly from ab initio calculations and enabling precision extrapolations beyond the reach of conventional schemes. 
This review focuses on model-space extrapolation methods developed for the no-core shell model and related many-body methods. 
We discuss machine learning extrapolation frameworks in comparison to conventional methods and assess their performance for energy spectra, radii, and electromagnetic observables, with particular emphasis on achievable precision and uncertainty estimates through statistical and correlation-based strategies.
These developments establish machine learning as an increasingly important component of the precision toolbox in ab initio nuclear theory, enhancing the reliability and predictive power of ab initio nuclear structure calculations.

\end{abstract}

\section{Introduction}

Over the past decade, machine learning (ML) has emerged as a versatile set of tools across most if not all fields of science, including nuclear physics. 
Initially popularized in areas such as image recognition and natural language processing, ML methodologies are now also contributing to fundamental challenges in nuclear theory by learning from experimental data across the nuclear chart to enhance the predictive capabilities of theoretical models~\cite{DoSchu15,NeuCa18,LaRe20,GaWa21,MuSp22,YuSo24}. 
The rapid growth of interest in ML methods is reflected in recent comprehensive reviews, which highlight the broad applicability of ML to nuclear data, structure calculations, reaction theory, and uncertainty quantification~\cite{BoAm22,HeLi23,CaCi19}.

Simultaneously, ab initio nuclear theory has entered a precision era. 
With the interplay of increased computational capabilities and conceptual advances for large-scale many-body simulations~\cite{Hergert20}, refined interaction models from chiral effective field theory (EFT)~\cite{EpEv20,MaSa24}, and sophisticated Bayesian schemes for uncertainty quantification and parameter estimation~\cite{FuKl15,MeFu19,SveEk23}, the precision of theoretical predictions is gradually approaching experimental precision.
This marks an important threshold for future refinement of interaction models, searches for beyond standard model physics, and guidance for the next generation of experiments.

A particularly interesting development at the intersection of both of these trends is the application of ML in ab initio nuclear structure calculations.
Although the data-driven nature of ML techniques seems to contradict the ab initio principles of controlled uncertainties and systematic improvability, they have become powerful supplemental tools for the computationally expensive large-scale ab initio many-body methods. 

Prominent examples include neural-network quantum states in quantum Monte Carlo frameworks, where they serve as representations for correlated many-body wave functions~\cite{LoCa26}, emulators based on Gaussian processes as efficient surrogate models for expensive large-scale calculations~\cite{KeHe23,BeMu26}, or Bayesian optimization for exploring high-dimensional parameter spaces in interaction calibration~\cite{EkFo19}. 
Besides their impressive successes, these ML applications generally do not increase the precision of theoretical predictions as they do not extend beyond accessible model spaces. 

Therefore, the topic central to the present review is the model-space extrapolation problem that poses a major limitation to the precision of nuclear structure calculations.
Many modern ab initio many-body methods rely on truncated basis expansions to render the nuclear many-body problem tractable~\cite{BaNa13,HaPa14,HeBo16,Soma20}. 
Recovering the full Hilbert space limit, ideally with quantified uncertainty, has long been a formidable challenge. 
Although, sophisticated renormalization group methods such as the similarity renormalization group (SRG) have accelerated the convergence rate of many-body calculations w.r.t.\ their model-space truncations~\cite{BoFu7,BoFu10,RoNe10,FuHe13,RoCa14}, exact solutions remain out of reach for all but the lightest nuclei.
In recent years, data-driven extrapolation schemes based on ML techniques have emerged as postprocessing tools for ab initio calculations that provide precise predictions of full-space observables with statistically robust uncertainties~\cite{NeVa19,JiHa19,MaSha24,KnoWo23,WoKno24,KnAg25}. 

Throughout this review, we survey modern ML extrapolation schemes alongside conventional methods and assess their predictive capabilities with a particular focus on uncertainty quantification and precision.
We further discuss the extent to which these ML tools enable high-precision predictions of different observables across a range of p-shell nuclei that allow for important quantitative insights into nuclear interaction models.
Finally, we conclude with a perspective on future opportunities and challenges for ML in high-precision ab initio nuclear theory.

\section{A brief introduction to artificial neural networks}

The ML approaches discussed in this work are based on artificial neural networks (ANNs). 
This section is meant to provide a short introduction to ANNs that covers only the most relevant basics required for understanding the following applications.
By no means should this be considered a fundamental or even remotely complete picture of the wide field of ANNs and we refer the interested reader to some introductory literature provided in Ref.~\cite{GoBe16}.

ANNs consist of neurons that are loosely modeled after their biological counterpart.
A single neuron, also called node, receives a set of inputs from which it computes an output.
Mathematically the neuron computes the output $y$ from the weighted sum of the received inputs $x_i$, adds a bias $b$, and evaluates an activation function $\sigma$ that typically introduces a non-linearity.
\begin{flalign}
    y = \sigma \qty(\sum_i x_i w_i + b)
\end{flalign}
The neurons in an ANN are organized in layers and, in all applications discussed in here, each layer is fully connected to the next one, i.e., each neuron in a given layer receives the outputs from all neurons in the previous layer as inputs and passes its output on to every neuron in the next layer.
The weights $w_i$ denote the respective strength of each connection.

ANNs are generalized function approximators that are used to model an unknown functional dependency between their inputs and outputs.
All weights and biases in the network are adjustable parameters that need to be constrained on data in a process known as training.
The ANNs in this work are trained using supervised training.
Based on a set of training data that comprises pairs of input data and known outputs, the parameters of an ANN are iteratively optimized using a statistical backpropagation algorithm~\cite{Rojas96} until the known outputs of the training data are reproduced to a desired accuracy.
Outside of the training data no additional knowledge about the form of the functional dependency is required.
Once trained the ANNs can be used to predict outputs for previously unseen inputs.

The quality of ANN predictions depends on a variety of different aspects including the quality of the training data, the topology of the ANN, choices of activation function and backpropagation algorithm, and many more. 
For the sake of brevity we refrain from going into more detail here, however, extensive discussion of the respective ANN setups can be found in the references provided for each application.

\section{The model-space extrapolation problem}

Ab initio many-body methods are explicitly designed to produce results that converge towards the exact solution in the full Hilbert space as the controlling truncations are gradually lifted.
Hence, when solved for a sequence of increasing model-space sizes a convergence pattern emerges. 
Despite tremendous progress in the past few decades, actual convergence is inaccessible for all but the smallest nuclear systems due to computational limits and will remain out of reach for most nuclei for the foreseeable future. 
This naturally raises the question on how to extrapolate results in truncated model spaces to the full Hilbert space.

The difficulty of this extrapolation problem depends strongly on the complexity of the convergence pattern, which itself can depend on various factors such as the observable under investigation, the choice of Hamiltonian, the utilized many-body method and its truncations, the underlying single-particle basis, and more.
While the concepts discussed here are generally applicable to any many-body method and truncation scheme, the discussion in this work is focused on the no-core shell model (NCSM) \cite{NaQu09,BaNa13} as one of the simplest yet most powerful ab initio many-body methods if convergence was practically achievable.
However, its reach is severely limited due to rapid model-space growth.
All NCSM calculations discussed in this work have been performed either with interaction families from chiral EFT with nucleon-nucleon and three-nucleon forces, or the adapted Daejeon16 interaction~\cite{ShiShi16}. 
These interactions are explicitly referred to as either EMN[$\Lambda$] for the non-local interaction developed by Entem, Machleidt, and Nosyk~\cite{EnMa19,HueVo20} or SMS[$\Lambda$] for the semi-local interaction introduced by the LENPIC collaboration~\cite{ReKre18,LENPIC21,LENPIC22}, where $\Lambda$ specifies the respective cutoff value in MeV.

In the NCSM, the convergence pattern is predominantly controlled by the choice of harmonic oscillator (HO) frequency \hw for the single-particle basis and the model-space truncation parameter \Nmax. 
The completeness of the HO basis is unaffected by the choice of \hw.
Therefore, the calculations become independent of \hw for $\Nmax\rightarrow\infty$ and all calculations converge towards the exact many-body solution.
\begin{figure}
     \centering
     \begin{subfigure}[b]{0.49\textwidth}
         \centering
         \includegraphics[width=\textwidth]{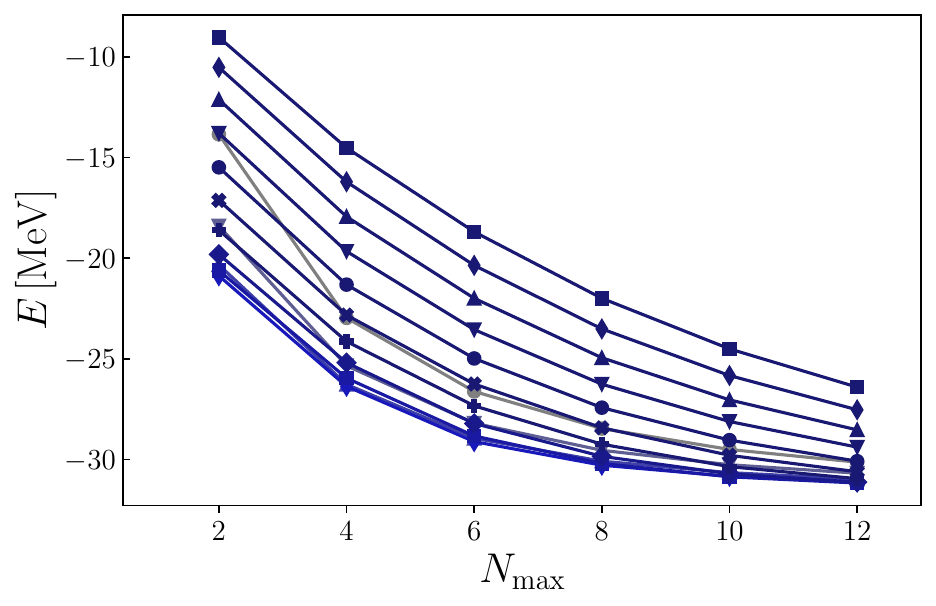}
     \end{subfigure}
     \hfill
     \begin{subfigure}[b]{0.49\textwidth}
         \centering
         \includegraphics[width=\textwidth]{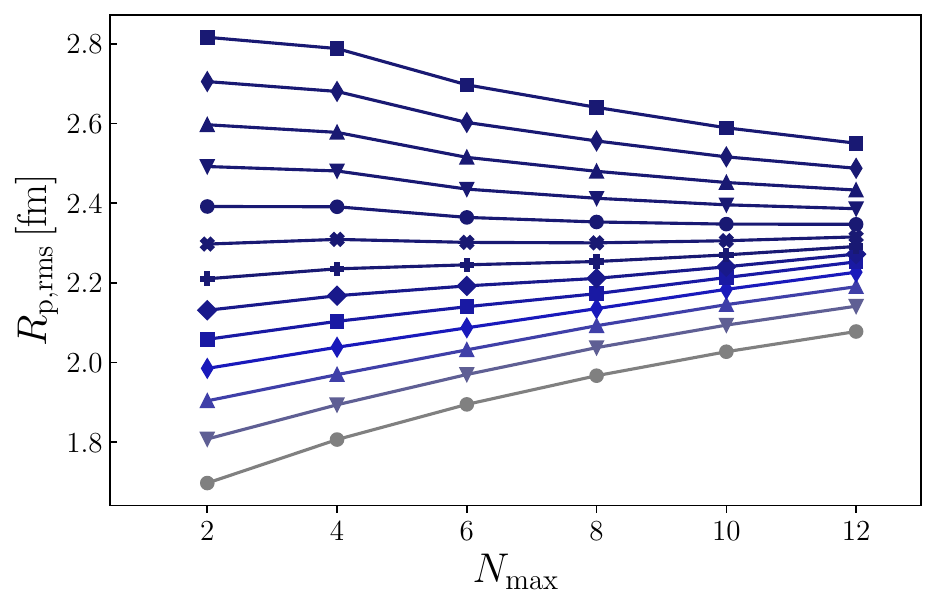}
     \end{subfigure}
     \caption{NCSM calculations from \cite{KnAg25} for the ground-state energy (left-hand panel) and point-proton radius (right-hand panel) of \elem{Li}{6} for an equidistant grid of oscillator lengths from $1.2$ to $2.4$~fm (grey to blue) for the EMN[500] interaction.}
     \label{fig:conv_patterns}
\end{figure}
Illustrations of the most common convergence patterns are shown in \cref{fig:conv_patterns}.
The energy convergence, constrained by the variational principle, only converges from above while the radius sequences can convergence from both sides.

\section{Modeling extrapolation functions}\label{sec:extrap_func}

The most intuitive way to address the extrapolation problem is to model a function that accurately describes the convergence pattern in the range of accessible model spaces.
This function can then be evaluated for truncation parameters equivalent or close to the full Hilbert space. 
In this section, we review three different ways of modeling these extrapolation functions based on heuristic considerations, physics-motivated reasoning, and data-driven machine learning concepts. 

\subsection{Heuristic approaches}

Traditionally, extrapolation functions have been modeled heuristically with varying degrees of success. 
The data is commonly described with polynomial or exponential functions~\cite{BoFu08,Roth09,MaVa09,LENPIC21}, where
the latter is particularly useful for ground-state energies in configuration interaction calculations, which obey the variational principle and, therefore, exhibit a monotonously decreasing convergence behavior as a function of \Nmax reminiscent of an exponential curve.
In such cases, a reasonably accurate extrapolation function at a fixed \hw is given by
\begin{flalign}
    E^\hw(\Nmax) = E^\hw_\infty + a \exp(-b\Nmax)\label{eq:exponential}
\end{flalign}
where \Nmax is the parameter controlling the model-space size and $E^\hw(\Nmax)$ is the energy obtained in the respective model space.
The parameters $a$, $b$, and $E^\hw_\infty$ need to be constrained by a sequence of calculations and $E^\hw_\infty$ yields the extrapolated result for the full Hilbert space.
The superscript \hw indicates that, in practice, $E^\hw_\infty$ retains some frequency dependence.
This dependence can be reduced by fitting a set of exponential equations 
\begin{equation}
\begin{split}
    E^{\hw_1}(\Nmax) &= E_\infty + a_1 \exp(-b_1\Nmax)\\
    E^{\hw_2}(\Nmax) &= E_\infty + a_2 \exp(-b_2\Nmax)\\
    &\;\;\vdots\\
    E^{\hw_n}(\Nmax) &= E_\infty + a_n \exp(-b_n\Nmax)
\end{split}\label{eq:ensemble}
\end{equation}
to multiple data sequences for different values of \hw simultaneously ensuring that all sequences share the same limit $E_\infty$. 
Put differently, we can think of the extrapolation function as being informed with the additional knowledge that all sequences converge towards the same limit. 
This `ensemble fit' is an interesting concept, which we will come back to later. 

The challenge with observables other than energies is the lack of constraints on the convergence patterns, which can virtually take on any shape.
Hence, there are no widely used extrapolation models although the extrapolation of radii has long been of particular interest and multiple avenues have been explored:
Radii exhibit a strong sensitivity to \hw one can exploit to fine tune the convergence rate, however, this is a rather exploratory than methodical process.
It has further been observed that if plotted over \hw the sequences for different \Nmax seem to cross at approximately the same value which has become known as the crossing point method~\cite{BoFu08} but some studies suggest a limited applicability of this heuristic estimate~\cite{CaMa14}.
Finally, for sufficiently large values of \hw the radius sequences adapt a mostly monotonously increasing pattern, which can again be modeled with an exponential ansatz.

In order to assign uncertainties to these extrapolations a variety of recipes has been developed, which mostly rely on multiple fits to some or all possible subsets of the available data (see, e.g., Ref.~\cite{LENPIC21}). 
Although this produces extrapolation estimates with uncertainties in a simple and rapid manner, the initial assumptions about the shape of the extrapolation function present a strong bias rendering the model inherently inaccurate.
Therefore, heuristic extrapolation schemes are rarely applicable beyond ground-state energies.

\subsection{Physics-motivated approaches}
Another class of extrapolation methods that has been widely applied to ground-state energies are so-called infrared (IR) extrapolations based on effective theories~\cite{FuHa12}.
This physics-motivated approach leverages the ultraviolet (UV) and IR cutoffs of the HO basis, which many methods including the NCSM are based on.
A correction to the energy at a given model-space size is derived from the insight that for long wavelengths the finite HO basis becomes indistinguishable from a spherical well with periodic boundary conditions.
An effective UV momentum $\Lambda_\mathrm{UV}$ and IR length scale $L_\mathrm{IR}$ can be expressed in terms of \Nmax and \hw.
For sufficiently large UV momenta, i.e., $\Lambda_\mathrm{UV}$ exceeds the momentum cutoff of the employed interaction, the calculation is considered UV converged and the remaining dependence on $L_\mathrm{IR}$ in leading order is given by
\begin{flalign}
    E(L_\mathrm{IR}) = E_\infty + c_1 \exp(-2c_2L_\mathrm{IR}).\label{eq:IR_energy}
\end{flalign}
We emphasize the similarity to \cref{eq:exponential}, however, the precise determination of $L_\mathrm{IR}(\Nmax,\hw)$ is non-trivial.
Several approximations, some adapted to specific many-body methods, as well as sub-leading corrections to \cref{eq:IR_energy} have been derived in the works led by Coon, Furnstahl, More, and Wendt~\cite{FuHa12,CoAve12,MoEk13,FuHa15,WeFo15}.
Analogously, corrections to the squared radius have been derived and read
\begin{flalign}
    \langle r^2\rangle(L_\mathrm{IR}) = \langle r^2\rangle_\infty \qty[1-c_0(2c_2L_\mathrm{IR})^3\exp(-2c_2L_\mathrm{IR})]
\end{flalign}
in leading order, where $c_2$ is the same as in \cref{eq:IR_energy} and typically determined in the energy fit. 
Although available, IR radius extrapolations are rarely employed in actual calculations. 

While these IR extrapolations are well founded in the underlying theoretical framework, their application is more complicated compared to the heuristic approaches and it can be difficult or computationally expensive to obtain enough UV-converged data as uncommonly large values of \hw far from the variational optimum are required.
Moreover, there is, as of now, no robust quantification procedure for the systematic uncertainties that goes beyond the arbitrary omission of data subsets in the fitting process. 

Another noteworthy approach for the correction of radii obtained in finite model spaces has recently been developed by Sun et al.~\cite{SuLe25}. 
They reformulate radii in terms of two-body densities, which are known to have exponential tails in the full-space solution.
However, the tails in truncated model spaces typically fall off faster than exponential due to the employed HO basis, resulting in an underestimation of radii, which are particularly sensitive to the tail of the density. 
The authors account for this shortcoming by identifying a range $r$ above which they replace the calculated tail by an exponential function.
Corrected radii are then obtained from the modified and re-normalized density.
While this is not an explicit model-space extrapolation method in the classical sense, it is a promising lead to address deficits caused by model-space truncations. 

\subsection{Machine learning approaches}

More recently, ML methods have been explored as extrapolation tools.
An interesting approach developed by Yoshida~\cite{Yoshida20} makes use of constrained Gaussian processes where results at large \Nmax are predicted from calculations in small model spaces in a Bayesian statistics framework.
The main advantage of this method is that it can operate on sparse data sets and inherently encompasses uncertainty estimates.
However, predictions of converged ground-state energies only become feasible with the inclusion of specific constraints, such as the monotonicity of the convergence pattern, and high precision requires the additional assumption of an exponential shape as in \cref{eq:exponential}.
The requirements of prior knowledge about the convergence pattern and observable-specific constraints severely complicate the extension to observables other than energies.
While the author mentions potential remedies for these issues, those avenues have not yet been explored.

We, therefore, shift our discussion to ANNs, which have been established as a more general data-driven alternative for the approximation of extrapolation functions.
On a fundamental level ANNs are general function approximators and therefore ideal candidates for modeling unknown functions such as the extrapolation functions we are interested in.
This idea has first been pioneered by Negoita et al.~\cite{NeLu18,NeVa19} and later been refined by Jiang et al.~\cite{JiHa19} and Shirokov and collaborators~\cite{MaSha24,ShaMa25}.
The idea behind this approach has a striking simplicity and can be broken down into three essential steps:
\begin{itemize}
    \item[1.] Produce data through many-body calculations for a variety of \Nmax and \hw.
    \item[2.] Train an ANN on this data until it has captured the desired functional dependency of the observable $O(\Nmax,\hw)$.
    \item[3.] Extrapolate by evaluating the ANN for very large values of \Nmax.
\end{itemize}
The main advantage of this ansatz over the previously discussed extrapolation schemes is that no prior assumption about the shape of the extrapolation function enters. 
Therefore, the ANN is not biased by the practitioners perception and it can be directly applied to any observable regardless of the convergence pattern.

\begin{figure}
     \centering
     \begin{subfigure}[b]{0.3\textwidth}
         \centering
         \includegraphics[width=\textwidth]{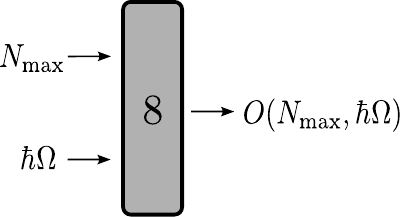}
         \caption{ISU ANN}
         \label{fig:topology_ISU}
     \end{subfigure}
     \hfill
     \begin{subfigure}[b]{0.32\textwidth}
         \centering
         \includegraphics[width=\textwidth]{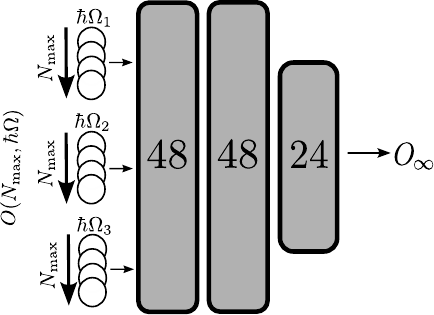}
         \caption{TUDa FSPN}
         \label{fig:topology_FSPN}
     \end{subfigure}
     \hfill
     \begin{subfigure}[b]{0.3\textwidth}
         \centering
         \includegraphics[width=\textwidth]{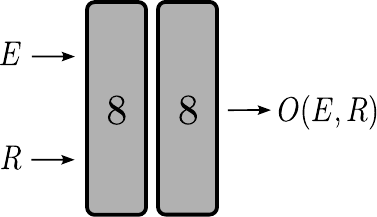}
         \caption{TUDa OTN}
         \label{fig:topology_OTN}
     \end{subfigure}
     \caption{Topologies of the different ANN applications. Gray rectangles denote hidden layers and the numbers indicate the hidden nodes. All layers are fully connected. See main text for details.}
\end{figure}

The ANN topology employed by Negoita et al., also referred to as ISU approach, 
is determined by the parameters of the extrapolation function. 
The input layer consists of two input nodes where \Nmax and \hw enter, while the output layer has a single neuron that yields the value of the observable for a given input.
The number of hidden layers and nodes may vary depending on the observable but was originally proposed as a single hidden layer with eight nodes.
A schematic illustration of the topology of an ISU ANN is given in \cref{fig:topology_ISU}.
The resulting ANN comes with 33 free parameters that need to be constrained during the training process.
A detailed description of the training process is given in Ref.~\cite{NeVa19}.

The amount of training data available for these applications is typically in the order of $\lesssim10^2$.
In principle, this data is sufficient to constrain the free parameters and learn the functional behavior, however, the generality of the ANN requires enough information in the training data to constrain them in a meaningful way.
In particular, the lack of assumptions about the shape of the extrapolation function
necessitates that the key features, i.e., convergence and \hw independence, must be learned solely from the training data.
The quality of this data is of essential importance as biases such as under- or overrepresented features will directly impact the performance of the ANN.
Moreover, a small set of training data often results in overfitting, i.e., the network describes the data accurately without learning the functional behavior, even with small ANN topologies.
Hence, those networks perform far better within the training domain than in the extrapolated region of interest.

These issues have been addressed by Jiang et al.~\cite{JiHa19} who were able to show that the training data set can be enlarged through interpolation techniques, thus, generating more high-quality training data and reducing overfitting.
They further showed the importance of a well-balanced data set to avoid pathological behaviors of the networks and reduce learned biases.
In addition, they provided first ANN extrapolations of coupled cluster calculations demonstrating the applicability to other many-body methods.

\setcounter{footnote}{0}

Nevertheless, an ANN trained this way often fails to reproduce the aforementioned key features.
In practical applications one therefore trains a multitude\footnote{Depending on the quality of the training data and the complexity of the convergence pattern `a multitude' can refer to anything from a few hundred up to a million trained ANNs.} of ANNs that all differ in their initial values for the free parameters, which are usually assigned randomly.
ANNs that fail to reproduce the key features are discarded and additional selection criteria may be applied, e.g., a threshold for the performance on selected testing data. 
Shirokov and collaborators have demonstrated how a sophisticated selection of ANNs motivated by the physics of nuclei can improve robustness and precision of the extrapolations~\cite{MaSha24,ShaMa25}.
However, these selection criteria should be chosen carefully since recent studies suggest that they might induce unwanted biases or lead to overly confident extrapolations~\cite{KnoLo25}.

Another advantage of training multiple ANNs is that we obtain a distribution of extrapolations instead of a single one.
From this distribution a final extrapolation along with an uncertainty can be extracted with statistical tools.
This uncertainty can be interpreted as a variation of the extrapolation function.

To date, ANN extrapolations based on the ISU approach have been successfully applied to energy spectra and radii in p-shell nuclei as well as hyperon separation energies in light hypernuclei~\cite{NeVa19,JiHa19,KnoLo25,Vidana23}. 
Some first applications to magnetic dipole (M1) and electric quadrupole (E2) moments can be found in Ref.~\cite{McCarty24}, yet their slow convergence w.r.t.\ \Nmax has proven particularly challenging for the networks.

Clearly, ANNs provide a powerful and versatile tool for the modeling of extrapolation functions. 
Their most striking feature is the applicability to any observable or many-body method due to the generality of the approach.
However, there are some caveats that need to be mentioned. 
First, the ANNs are only applicable where sufficient training data can be obtained.
Especially in situations with little data or data of low quality the training of the ANNs becomes inefficient as the majority of trained networks is being discarded.
Secondly, both, training data and ANNs have to be obtained for each combination of Hamiltonian, observable, and eigenstate.
Thus, a large number of many-body calculations as well as the whole training procedure need to be run for every application.
Finally, and most importantly, the model space problem remains an extrapolation problem, which are notoriously difficult for ANNs since they are explicitly designed as interpolators.
Hence, extrapolating far beyond the available training domain will always come with uncontrollable uncertainties.

\section{Predicting full-space solutions}

An alternative ML approach to the model-space extrapolation problem has been developed at TU Darmstadt and is therefore known as TUDa approach~\cite{KnoWo23,WoKno24}.
Instead of modeling an extrapolation function as discussed in \cref{sec:extrap_func}, this approach aims at directly predicting the solution in the full Hilbert space from calculations in truncated model spaces. 
Hence, the ANN is tasked to learn the actual convergence pattern of the respective observable.
Specifically, the converged value $O_\infty$ is predicted from a set of twelve values $O(\Nmax,\hw)$ calculated at four consecutive \Nmax steps for three different \hw.
This strategy mirrors an experienced human practitioner estimating the converged result from looking at the convergence pattern.
For distinctness from other ANN architectures we will refer to this type of ANN as full-space prediction network (FSPN).

Conceptually, the TUDa approach compares to the ISU approach in a similar manner as the ensemble fit in \cref{eq:ensemble} to the single exponential fit in \cref{eq:exponential}.
The FSPN is informed with data for multiple values of \hw and the information that all sequences converge towards the same limit is explicitly encoded in the networks topology.
As illustrated in \cref{fig:topology_FSPN} the input layer consists of twelve nodes for the data spanning the convergence pattern and the output layer contains a single node that yields the converged value.
This is a more complex task compared to the ISU approach, hence, a larger ANN is needed with three hidden layers consisting of 48, 48, and 24 nodes each.
This amounts to over 4000 adjustable parameters.

Training these parameters requires a large set of training data.
In order to predict the full-space solution the training data needs to be fully converged.
Such calculations are only computationally feasible for few-body systems. 
Hence, the FSPN is trained on a large set of NCSM data for \elem{H}{2}, \elem{H}{3}, and \elem{He}{4} for 36 different SRG evolved Hamiltonians and seven HO frequencies in an \hw window from 12 to 32~MeV resulting in more than 350,000 data samples.
The main assumption that enters here is that the observable-specific convergence pattern is very similar in these few-body systems and in p-shell and heavier nuclei.
More precisely, the variation in the convergence pattern from different Hamiltonians is expected to be larger than from the choice of nucleus, especially since the training data covers \elem{H}{2} as a loosely bound and spatially extended nucleus as well as the strongly bound and compact \elem{He}{4}. 
Normalizing the input data further eliminates the remaining nucleus-specific scales thereby increasing the validity of this assumption~\cite{WoKno24}.

The major difference to the previously discussed extrapolation approaches is that the problem is effectively converted into an interpolation task.
The FSPN does no longer extrapolate beyond the training domain but interpolate between the different shapes of convergence patterns that are included in the training data, thus playing to the strengths of ANNs.
This also means that, once trained, the same FSPN can be used to predict full-space solutions of the respective observable for all nuclei, eigenstates, and Hamiltonians as long as four consecutive model-space sizes are accessible.
It has further been shown that even hypernuclei can be extrapolated using the same FSPN despite the total absence of hypernuclear calculations in the training data~\cite{KnoRo23}.
Thanks to this universality, trained FSPNs can be stored and are readily available and directly applicable on demand without retraining, making this approach computationally efficient.  

Analogously to the ISU approach, uncertainties are estimated through the use of multiple FSPNs.
In addition, all possible input samples, i.e., subsets of twelve data points for four consecutive \Nmax and three values of \hw and their permutations, that can be constructed from the data are passed through the FSPNs, thereby further increasing the number of predictions and encompassing the sampling as another source of uncertainty.

\begin{figure}[t]
     \centering
     \begin{subfigure}[b]{0.49\textwidth}
         \centering
         \includegraphics[width=\textwidth]{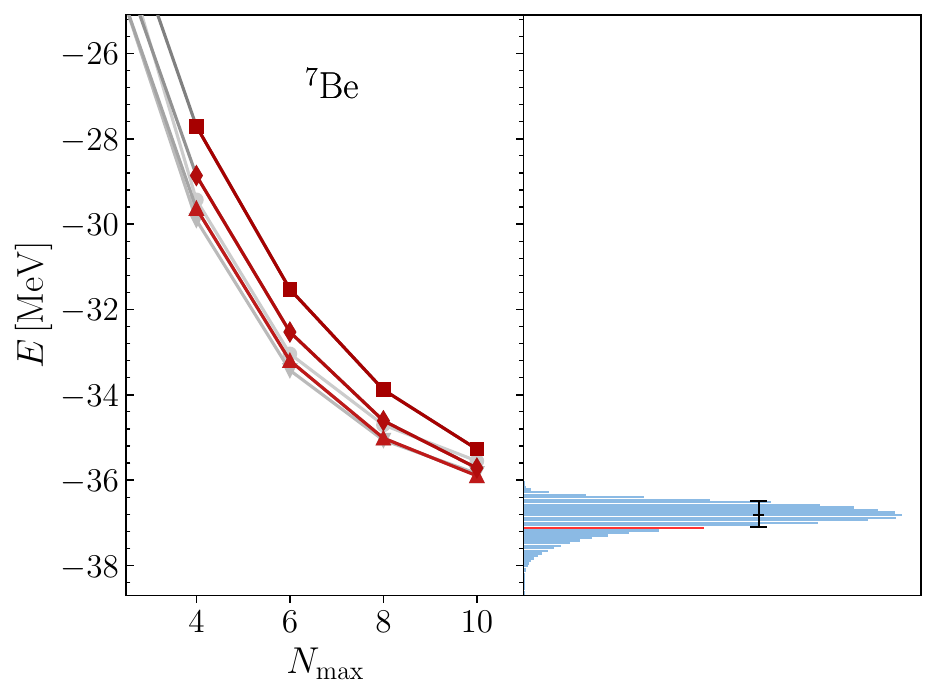}
     \end{subfigure}
     \hfill
     \begin{subfigure}[b]{0.49\textwidth}
         \centering
         \includegraphics[width=\textwidth]{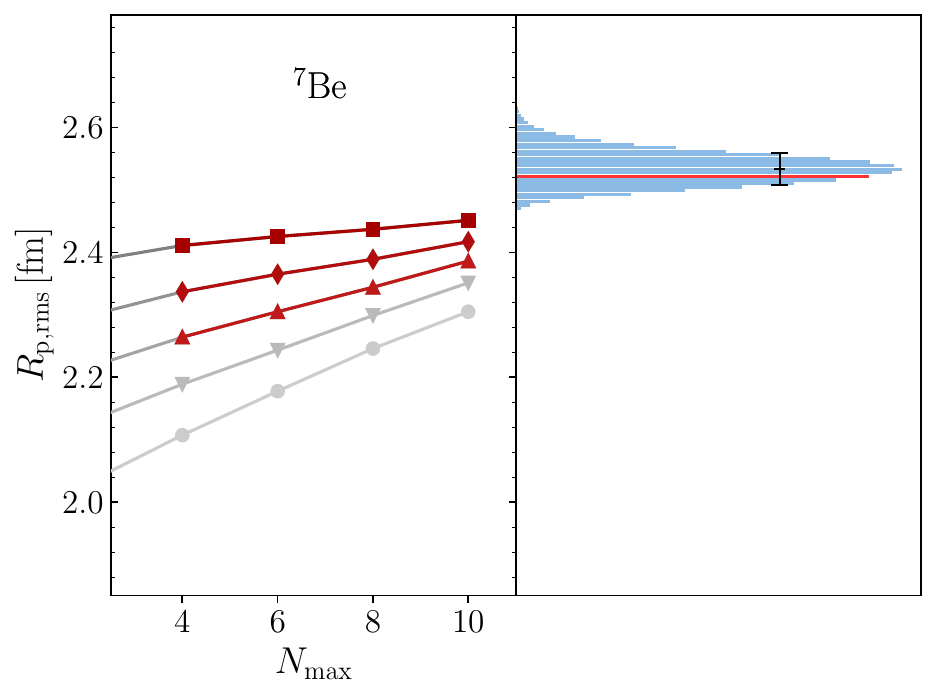}
     \end{subfigure}
     \caption{Examples of FSPN predictions for the ground-state energy and point-proton radius in \elem{Be}{7} from \cite{KnAg25}. The data and bars colored in red indicate a single input sample and the corresponding prediction for that sample.}
     \label{fig:FSPN_results}
\end{figure}

The TUDa approach has successfully been applied to energies~\cite{KnoWo23} and radii~\cite{WoKno24} and examples for ground-state energies and point-proton radii are presented in \cref{fig:FSPN_results}. 
These examples illustrate how a single prediction is obtained from a specific subset of the available data and the resulting distributions emerge from the combination of all possible predictions. 
The radius extrapolations additionally benefit from limiting the data to sequences converging from below, thus, simplifying the convergence pattern.

A comparison between the ISU and TUDa ML approaches and the heuristic and IR extrapolation schemes has been conducted in a benchmark study of Li isotopes~\cite{KnoLo25}.
While all methods are mostly in agreement with each other when evaluated on sufficiently converged data, the FSPNs exceed the capabilities of the other extrapolation schemes in smaller model spaces, yielding precise and consistent predictions from very limited data already.
Although this establishes ML approaches as the preferred extrapolation strategy, some inconsistencies between the TUDa and ISU approaches have emerged. 
The authors have traced this back to overconfident uncertainty estimates in the ISU approach, however, this certainly requires more investigations in the future.

\subsection{Derived observables}\label{sec:diff_obs}

Ground-state properties are the most commonly studied features of nuclei, however, there are other interesting observables that are technically derived from other properties.
Some observables such as excitation energies are defined as differences between the same observable for two eigenstates of a nucleus, while others like separation energies are determined from differences between two nuclei.
Another important occurrence of derived observables arises in direct comparison with experiments, where e.g.\ radii of neighboring isotopes are in many cases more accessible, i.e., measurable with higher precision, as isotope shifts than in absolute terms.

\begin{figure}
    \centering
    \includegraphics{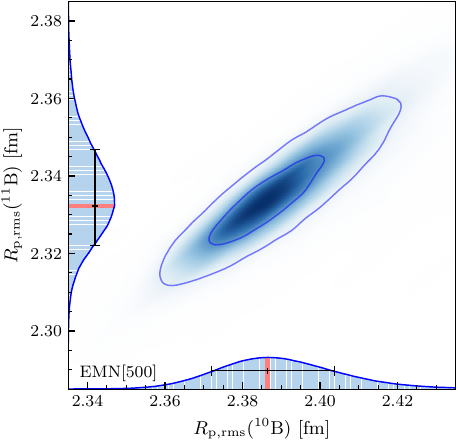}
    \caption{Correlation between the FSPN predictions for point-proton rms radii of the neighboring isotopes \elem{B}{10} and \elem{B}{11} from \cite{WoGe25} with their one dimensional projections. Inner and outer contours correspond to the $1\sigma$ and $2\sigma$ intervals of the projections. This correlation is leveraged to increase the precision of the isotope shift.}
    \label{fig:FSPN_diff_results}
\end{figure}

The convergence patterns for such derived observables often appear irregular compared to their `parent observables' and exhibit a much larger variety.
This impedes the training of FSPNs as no general shape of the convergence pattern can be identified and captured.
Instead of constructing a new set of ANNs the two components of the derived observable can be extrapolated individually with the existing FSPNs.
However, combining the uncertainties of the two components into an uncertainty for the derived observable is non-trivial since both components are typically strongly correlated as demonstrated in \cref{fig:FSPN_diff_results}.
Wolfgruber et al. have shown that these correlations can be exploited through sample-wise subtraction of the distributions~\cite{WoKno24,WoGe25}.
In this process, the input samples that share the same values for \hw and \Nmax are extrapolated separately, then subtracted, and the resulting value marks a single prediction for the derived observable.
Continuing for all pairs of input samples yields a narrow distribution of predictions that incorporates the correlations between the components.
This enables the prediction of full-space derived observables with increased precision.
Analogously, this technique has successfully been applied to predict hyperon separation energies in p-shell hypernuclei~\cite{KnoRo23}.

\subsection{Data scarcity for electromagnetic observables}

While the FSPN approach is, in principle, applicable to any observable, recent applications to electromagnetic observables show that insufficient training data quickly becomes a bottleneck for this method.
Many electromagnetic properties vanish in some of the few-body systems required for training, resulting in very limited amounts of training data with reduced diversity in convergence patterns. 
The work in Ref.~\cite{KnAg25} explores FSPN extrapolations of E2 moments.
Among the training nuclei only the deuteron exhibits a non-zero E2 moment, thus, the training data is limited to a single nucleus.
Nevertheless, FSPNs can still be constructed from this limited data and provide reliable predictions although they are more prone to nucleus-specific biases.

Such reduced data FSPNs are feasible for very few other electromagnetic properties, e.g.\ the magnetic dipole moment, however, this approach eventually becomes inapplicable for higher order electromagnetic observables.
Fortunately, there is a way to work around this limitation by using yet another ML tool to exploit correlations, thereby restoring access to the full suite of observables.

\section{Machine learning for correlated observables}

It is well-known that the convergence patterns of certain nuclear observables are highly correlated with each other.
Some observables such as the energies of different eigenstates in the same nucleus or radii in neighboring isotopes are clearly correlated due to similarities in the nuclear systems and we have already discussed in \cref{sec:diff_obs} how considering those yields large gains in precision.
Other correlations become apparent on the operator level.
A prime example is the E2 moment, which strongly correlates with the mean squared point-proton radius due to the $r^2$ dependence of the operator.
Caprio et al. have demonstrated that combining those into dimensionless quantities significantly accelerates model-space convergence and reduces frequency dependence~\cite{CaFa22,CaMa25}.
Hence, the correlations must already be present in comparably small model spaces.

These correlations can also be leveraged in the context of the model-space extrapolation problem through observable transcoder networks (OTNs) recently introduced in Ref.~\cite{KnAg25}.
As discussed so far, energies and radii can be reliably extrapolated to the full Hilbert space, while extrapolating electromagnetic observables in a similar manner remains challenging.
Hence, instead of directly predicting the full-space solution for electromagnetic observables, they can be obtained from extrapolated energies and radii through an observable transcoder function $O(E,R)$ that connects an observable $O$ to the energy and radius of the nucleus.
Although the shape of this function is unknown, it can once more be modeled using an ANN referred to as OTN.

The network's topology, as illustrated in \cref{fig:topology_OTN}, contains two input nodes for energy and radius respectively, two hidden layers with eight neurons each, and a single output node yielding the prediction for the observable of interest.
Since the correlations between these observables converge rather quickly with model-space size, the OTN can be trained on NCSM data from truncated model spaces.
Afterwards, a prediction for the full-space electromagnetic observable is obtained from full-space predictions of energy and radius, which, in turn, are accessible via any of the previously discussed extrapolation schemes.

It is conceptually interesting to note that this approach is partially an interpolation and partially an extrapolation problem. 
Energy and radius serve as a proxy for \Nmax and \hw, thus, reparameterizing the model-space convergence of observable $O$.
Due to its monotonicity the energy resembles a measure of convergence, while the radius indicates the length scale of the nucleus that is closely tied to \hw.
Therefore, a wide range of radii can be sampled in the training data through variation of \hw, which allows for interpolation in this dimension.
The energy dimension on the other hand requires an extrapolation as the full-space energy can only be approached from one direction. 
In this regard, the OTN is similar to an ISU ANN.
However, instead of the far extrapolation in \Nmax required in the ISU approach, the energy difference accounted for in the extrapolation of the OTN is typically much smaller than the energy range spanned by the training data.

In order to assign meaningful uncertainties to observables extrapolated with the OTN approach, two sources of uncertainty need to be considered.
The first one is the uncertainty of the OTN itself.
Analogous to all other ANN-based methods this uncertainty is quantified by using multiple OTNs. 
This mostly encompasses the uncertainty for the small extrapolation in the energy direction.
The second source are the individual uncertainties of the input energy and radius that need to be propagated to the transcoded observable.
If these inputs are available as distributions of values, e.g., as obtained from FSPNs, the uncertainties can be incorporated by sampling these distributions and evaluating the OTNs for each sample.
Otherwise, one could treat input and uncertainty as a normal distribution or similar in order to draw samples.
It turns out that the correlations can be modeled with such high fidelity that the limiting factor for the precision of the OTN prediction is the precision of the radius extrapolation.
Therefore, the OTNs yield predictions for E2 moments with unprecedented precision especially when considering the slow rate of convergence and strong \hw dependency~\cite{KnAg25}.

\begin{figure}
     \centering
     \includegraphics[width=.7\textwidth]{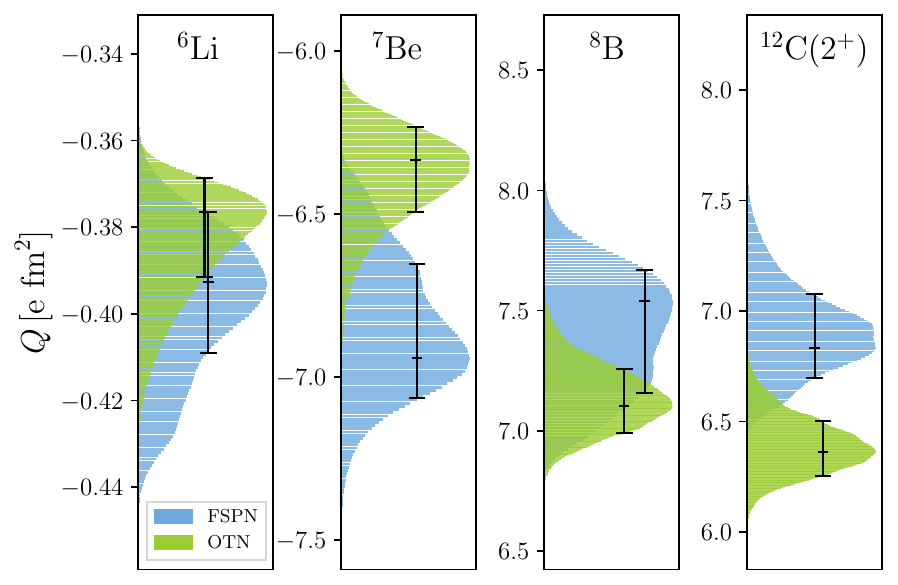}
     \caption{Comparison of FSPN and OTN predictions of E2 moments from~\cite{KnAg25} obtained with the EMN[500] interaction. Black error bars indicate the most probable values along with 68\% uncertainty intervals.}
     \label{fig:E2Mom_results}
\end{figure}
The E2 moment is an interesting observable that allows for the comparison of the OTN approach to the FSPN approach as depicted in \cref{fig:E2Mom_results}. 
Both methods yield predictions with similar precision and while the respective histograms all overlap to some degree the FSPN predictions tend towards larger absolute values and the associated histograms have a more irregular structure.
Both of these aspects can likely be attributed to the scarcity of training data for the FSPN approach, making the OTN ansatz favorable.

Being a very new approach, the OTN method has, to date, only been applied to the E2 moment.
However, it provides access to all other observables computable within the NCSM since energy and radius provide a complete reparametrization of the model space.
Moreover, the correlations between the observables are universal, i.e., independent of the employed many-body method.
This allows for cross-method applications such as OTNs being trained on NCSM data but evaluated with energies and radii from precise medium-mass many-body methods, thus, bringing nuclei beyond the p-shell into perspective.

\section{High-precision predictions with combined uncertainties}

In the preceding discussion, we have seen that ML-based extrapolation schemes allow for a statistical quantification of uncertainties from model-space truncations.
However, in order to assess the precision of a theoretical prediction in a meaningful way reliable estimates of all relevant sources of uncertainty are ultimately required.
While they are often assessed and compared individually, the combined treatment of uncertainties remains a formidable task.

The dominant source of uncertainty in modern ab initio calculations up to p-shell nuclei is typically the truncation of the chiral expansion in interaction models from chiral EFT~\cite{MaLe23}.
Its rather slow convergence raises the need to include higher orders of the expansion.
To date, even the most refined families of interactions are only consistently available up to third or fourth order and many-body forces beyond three-nucleon contributions are impractical for now~\cite{MaSa24}.
In addition, the low-energy constants (LECs) that occur as free parameters in the expansion need to be constrained on experimental data, where the choice of data and fitting procedure gives rise to another yet related source of theoretical uncertainty.

The determination of LECs and their uncertainties is nowadays often tackled with statistical methods such as Bayesian inference~\cite{SveEk23}.
These methods require large numbers of calculations to provide statistically robust results.
This has recently become feasible with the development of powerful surrogate models also known as emulators that are based on eigenvector continuation or ML techniques, particularly Gaussian processes~\cite{KoeEk20,DriMe23}. 
Emulators allow for the study of parameter dependencies in the interaction model through very efficient probing of high-dimensional parameter spaces.
This way, parameter uncertainties in the interaction can be propagated through computationally expensive many-body solvers at greatly reduced cost.
As a discussion of these methods in more detail goes beyond the scope of this review, the interested reader is referred to Refs.~\cite{EkFo23,DuEk24} and the references therein for further reading.

\begin{figure}
     \centering
     \begin{subfigure}[b]{0.62\textwidth}
         \centering
         \includegraphics[width=\textwidth]{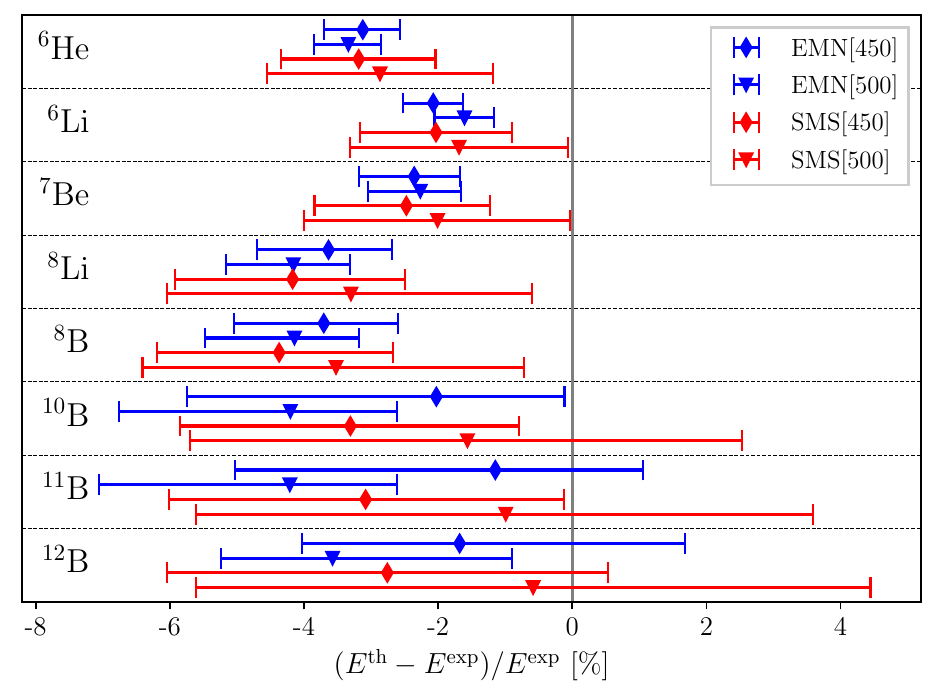}
     \end{subfigure}
     \hfill
     \begin{subfigure}[b]{0.62\textwidth}
         \centering
         \includegraphics[width=\textwidth]{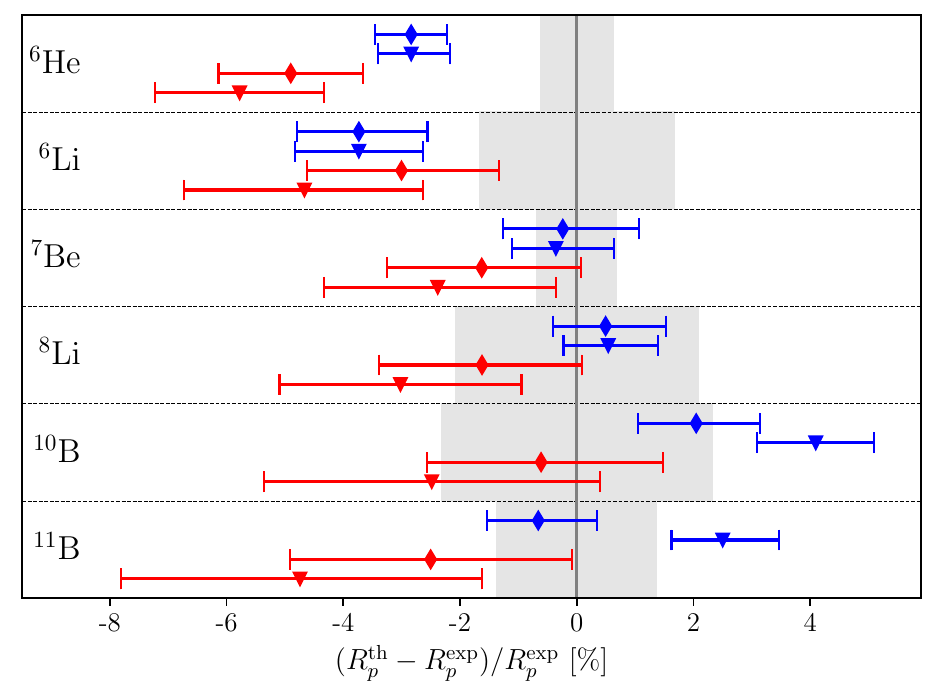}
     \end{subfigure}
     \hfill
     \begin{subfigure}[b]{0.62\textwidth}
         \centering
         \includegraphics[width=\textwidth]{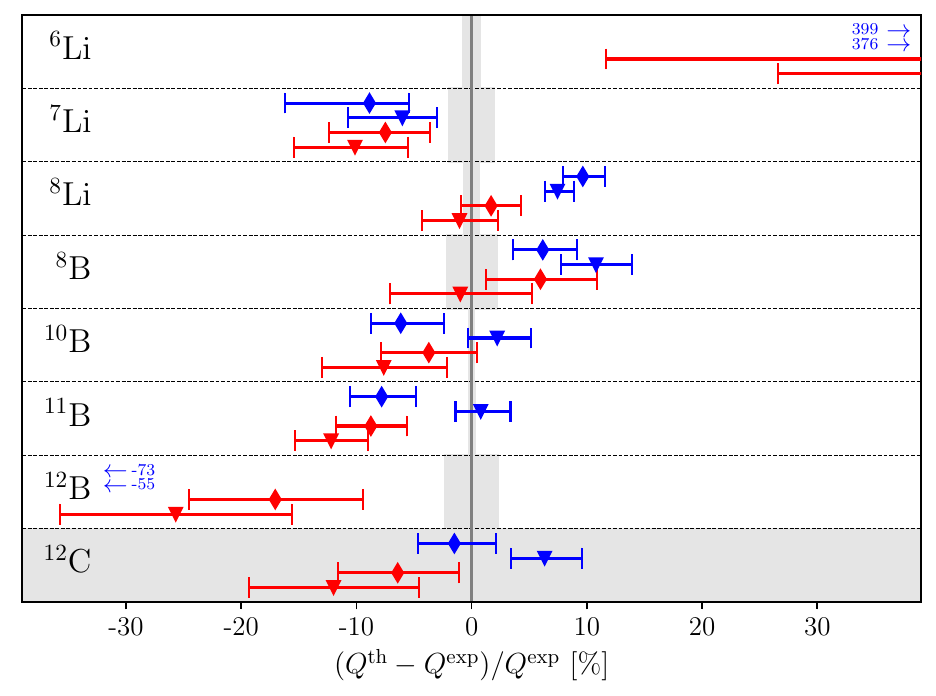}
     \end{subfigure}
     \caption{Relative deviation between theoretical prediction and experiment for ground-state energies (top panel), point-proton radii (center panel), and electric quadrupole moments (bottom panel) for a range of p-shell nuclei collected from Refs.~\cite{WoGe25,KnAg25} with experimental values taken from Refs.~\cite{WaHua21,AnMa13,Stone16}. Different colors indicate different families of chiral interactions.}
     \label{fig:precision_results}
\end{figure}

Estimates for chiral truncation errors are based on the perturbative nature of the chiral expansion.
Hence, some early heuristic approaches estimate the truncation error from power-counting assumptions in an order-by-order analysis~\cite{EpKr15,EpEv20}. 
A more sophisticated approach based on Bayesian statistics but following the same power-counting assumptions has been pioneered by Furnstahl et al.~\cite{FuKl15} and later extended and refined by the BUQEYE collaboration~\cite{MeWe17,MeFu19}.
Here, the omitted higher-order terms are modeled as random variables with specified priors.
These priors are then updated with information from results obtained at accessible chiral orders.
For the observables discussed here, the BUQEYE pointwise model~\cite{MeFu19} provides an easily applicable procedure for the estimation of truncation uncertainties, resulting in a Student-$t$ probability distribution that quantifies credibility intervals around the highest-order value. 

In order to provide combined uncertainty estimates that include interaction uncertainties as well as many-body uncertainties, Refs.~\cite{WoGe25,KnAg25} propose to calculate the convolution of the distribution of predictions obtained from the previously discussed ML extrapolation schemes with the BUQEYE Student-$t$ distribution that resembles the interaction uncertainty.
The resulting distribution then contains information of both errors and allows for the statistical extraction of combined uncertainties.

Based on this combined uncertainty quantification for ML extrapolation schemes and interaction uncertainties a comparison with experiment becomes meaningful.
\cref{fig:precision_results} presents a collection of results for energies, radii, and E2 moments over a range of p-shell nuclei obtained within the FSPN and OTN frameworks for different interaction models from chiral EFT.
Energies are consistent across interactions, yet, systematically underestimate experiment.
Radii and E2 moments, in turn, are less consistent across interactions but some nuclei are described very accurately.
Even with both major sources of uncertainty considered, these calculations are precise enough to allow for a quantitative assessment of different interaction models, a crucial step towards a more refined understanding of the strong interaction in nuclei.

\section{Summary and outlook}

This review set out to provide a comprehensive overview of ML approaches that supplement ab initio calculations of nuclei with particular emphasis on ML extrapolation methods and precision aspects.
We conclude with a brief summary of the main takeaways and an outlook on current challenges and future opportunities.

ML approaches, and ANNs in particular, are versatile tools that allow for data-driven and, therefore, largely unbiased model-space extrapolation.
In contrast to conventional extrapolation schemes, these approaches provide meaningful uncertainty estimates and achieve consistency and precision exceeding those of established methods.
In fact, the theoretical predictions are sufficiently precise to enable quantitative analyses and comparison of different interaction models, which is an important requirement to deepen our understanding of EFTs for the strong interaction.

A current limiting factor on the precision frontier is the radius extrapolation due to its sensitivity to the long-range part of the wavefunction.
Here, new ideas that concentrate on the tails of wavefunctions may yield further improvements.

Another favorable feature of ANN extrapolations is their generality. 
By design, they are applicable to every observable, thus, providing the first systematic extrapolation methods for observables beyond energies and radii.
Especially OTNs trained on ab initio many-body data not only enable predictions of electromagnetic observables with unprecedented precision, they also open up an exciting pathway towards future applications that combine the strengths of multiple many-body methods.

Overall, ML extrapolation methods in ab initio nuclear theory are still at an early stage of development and additional work is required to fully understand remaining inconsistencies between the different approaches.
Plenty of avenues towards other observables such as electromagnetic transitions and applications to other many-body methods are yet to be explored.
Nevertheless, they already yield promising perspectives for high-precision theory predictions across a wide range of observables and have, thus, become a valuable asset in the precision toolbox. 

\ack{I thank Tobias Wolfgruber for his support with data and figures and his remarks on the manuscript.
I further thank Robert Roth for helpful advice and discussions.}

\bibliographystyle{iopart-num-mod}
\bibliography{references}

\end{document}